\documentclass[preprint, 12pt]{elsarticle}

\usepackage{amsmath,amssymb}
\usepackage{graphicx, color}
\usepackage{subfig}
\usepackage[colorlinks=true] {hyperref}
\journal{Commun Nonlinear Sci Numer Simulat}

\begin{document}

\begin{frontmatter}

\title{Front propagation into unstable states in discrete media}

\author[pucv]{K. Alfaro-Bittner\corref{cor}}
\ead{kapaalbi@gmail.com}
\author[uch]{M. G. Clerc}
\author[pucv]{M. Garc\'ia-\~Nustes}
\author[pucv]{R. G. Rojas}

\cortext[cor]{Corresponding author}

\address[pucv]{Instituto de F\'isica, Pontificia Universidad Cat\'olica de Valpara\'iso, 
Casilla 4059, Valpara\'iso, Chile}

\address[uch]{Departamento de F\'isica, Facultad de Ciencias F\'isicas y Matem\'aticas, 
Universidad de Chile, Casilla 487-3, Santiago, Chile}

\begin{abstract}

Non-equilibrium  dissipative systems usually exhibit multistability, leading to the
presence of propagative domain between steady states. We investigate the front propagation
into an unstable state in discrete media. Based on a paradigmatic model of 
coupled chain of oscillators and populations dynamics, we calculate  analytically 
the average speed of  these fronts 
and  characterize numerically the oscillatory front propagation. 
We reveal that different parts of the front oscillate with the same frequency 
but with different amplitude. To describe this latter phenomenon we generalize the notion of 
the Peierls-Nabarro potential, achieving an effective continuous description of the discreteness effect.

\end{abstract}


\end{frontmatter}

\section{Introduction}
\label{intro}

Macroscopic systems under the influence of injection and dissipation 
energy, momenta, or matter  usually exhibit 
coexistence of different states---this feature is usually denominated 
multistability \cite{Prigogine,Pismen}. 
Inhomogeneous initial conditions, usually caused by the inherent fluctuations, 
generate  domains that are separated by their respective interfaces. 
These interfaces are known as  front solutions, interfaces, domain walls or wavefronts
\cite{Pismen,Cross1993,CrossBook},
depending on the physical context where they are considered. 
Interfaces  between these metastable states appear in the form of propagating 
fronts and give rise to rich spatiotemporal dynamics  \cite{Pomeau1986,Langer1980,ColletEckman}.  
Front dynamics occurs in systems as different as walls separating magnetic 
domains \cite{Eschenfelder}, directed solidification processes \cite{Langer1980}, 
nonlinear optical systems \cite{Clerc2001,Gomila2001,Clerc2004,Residori2005},  
oscillating chemical reactions \cite{Petrov1997}, fluidized granular media 
\cite{Aranson2008,Douady1989,Moon2003,Macias2013},  
and population dynamics \cite{Murray1989,Fisher1937,Clerc2005}, to mention a few. 

In one spatial dimension---from the point of view of dynamical systems theory---a 
front is a nonlinear solution that is identified in the co-moving frame system as a 
heteroclinic orbit linking two steady states \cite{Saarloos1992,Coullet2002}. 
The evolution of front solutions can be regarded as  a particle-type one, i.e., 
they can be characterized by a set of continuous parameters such as position,  
core width and so forth.
The front dynamics depends on the nature of the steady states that are 
connected.  In the case of a front connecting two stable uniform states,  
a variational system 
tends to minimize   
its energy or Lyapunov functional. Thus, the front solution always propagates with a 
well defined unique speed towards  
the less energetically favorable 
steady state. There is only one point in the parameter space for which the front 
is motionless, the Maxwell point.  In   systems 
with  only local interaction between adjacent neighbors---a discrete medium---the front solutions 
persist \cite{Fath,ClercEliasRojas}.
Once more, the most favorable state invades the less favorable one, being now
the speed oscillatory. Actually, there is a region of the parameter space, close 
to the Maxwell point---the pinning range---where the fronts are motionless \cite{ClercEliasRojas}.
These  properties can be explained by considering a potential for the front position,
the Peierls-Nabarro barrier \cite{FrenkelKontorova}, which it is a 
result of the discreteness of the system.

The former scenario changes drastically for a front connecting  a stable and 
an unstable state, usually called  
Fisher-Kolmogorov-Petrosvky-Piskunov (FKPP) front 
\cite{Murray1989,FKP1937,Saarloos2003}.  FKPP fronts 
have been observed in auto-catalytic chemical reaction \cite{Luther}, 
Taylor-Couette instability \cite{Ahlers1983}, Rayleigh-Benard 
experiments \cite{Fineberg1987}, pearling and pinching on the propagating 
Rayleigh instability \cite{Powers1997}, spinodal decomposition in polymer mixtures 
\cite{Langer1992}, and liquid crystal light valves with optical feedback \cite{Clerc2004}.
  One of the features of these fronts is that  their speed is determined by initial conditions. 
 When initial conditions are bounded, 
after a transient period, two counter propagative fronts emerge with the minimum 
asymptotic speed \cite{Murray1989,FKP1937,Saarloos2003}.  In discrete media, 
FKPP fronts also persist exhibiting an oscillatory behavior, however, the pinning 
phenomenon no longer exists. 
Beyond that, there is few understanding about their general behavior in 
discrete systems in our knowledge.

The aim of this article is to investigate, theoretically and numerically, the FKPP front propagation
 in discrete media. Based on paradigmatic models: the dissipative Frenkel-Kontorova  
 and the discrete Fisher-Kolmogorov-Petrosvky-Piskunov equation. We determine 
analytically the  minimum mean speed of  FKPP fronts as a function of the medium discreteness.  
Numerically we characterize the oscillatory front propagation. 
We reveal that the front behaves as an extended object. Different 
parts of the front oscillate with the same 
frequency but with different amplitude. To describe 
this phenomenon we generalize the notion of the  Peierls-Nabarro potential, which
allows us to have an effective continuous description of the discreteness effect.

\section{Chain of dissipative coupled pendula}

Let us consider a chain of dissipative coupled pendula, known as 
{\it the dissipative Frenkel-Kontorova model},
\begin{equation}
\ddot{\theta}_i=-\omega^2\sin\theta_i-\mu \dot{\theta}_i+
\frac{\theta_{i+1}-2 \theta_i+\theta_{i-1}}{dx^2},
\label{Eq-ChainPendulum}
\end{equation}
where ${\theta}_i(t)$  is the angle formed by the pendulum
and the vertical axis in the $i$-position at time t, $i$ is the index label 
the $i$-th pendulum, 
$\omega$ is the pendulum natural
frequency, $\mu$ accounts for the damping  coefficient, and $dx$ stands for 
the interaction between adjacent pendulums.
This last parameter controls the degree of discreteness of the system.
When $dx \to 0$ the system describes front propagation in a continuous medium, 
the dissipative sine-Gordon equation.  
In the conservative or Hamiltonian limit, $\mu=0$, the above equation
is known as the Frenkel-Kontorova model,
which describes the dynamics of a chain of particles interacting
with the nearest neighbors in the presence of an external periodic potential.
The Frenkel-Kontorova model, Eq. (\ref{Eq-ChainPendulum}), is a paradigmatic 
model with application to several physical contexts. It has been used to describe the 
dynamics of atoms and
atom layers adsorbed on crystals surfaces,  incommensurate phase in dielectric, domain wall 
in magnetic domain,  fluxon  in  Josephson transmission lines, 
rotational motion of the DNA bases,
and plastic deformations in metals 
(see textbook \cite{FrenkelKontorova} and references therein).

Note that equation~(\ref{Eq-ChainPendulum}) can rewrite in the following manner
\begin{equation}
\mu\dot{\theta}_i=-\frac{\delta F}{\delta \theta_i},
\label{Eq-FunctionalEqPendulum}
\end{equation}
where the Lyapunov functional $F$ has the form
\begin{equation}
F\equiv \sum_{i=0}^N \left[\frac{\dot{\theta}_i^2}{2}-\omega^2 \cos\theta_i
+\frac{(\theta_{i+1}-\theta_{i+1})^2}{2 dx^2} \right].
\label{Eq-FunctionalPendulum}
\end{equation}
Hence, the dynamics of Eq.~(\ref{Eq-ChainPendulum}) is characterized by 
the minimization of functional $F$ when
$\mu \neq 0$.

\begin{figure}[tb]
\centering
\includegraphics[width=0.9 \columnwidth]{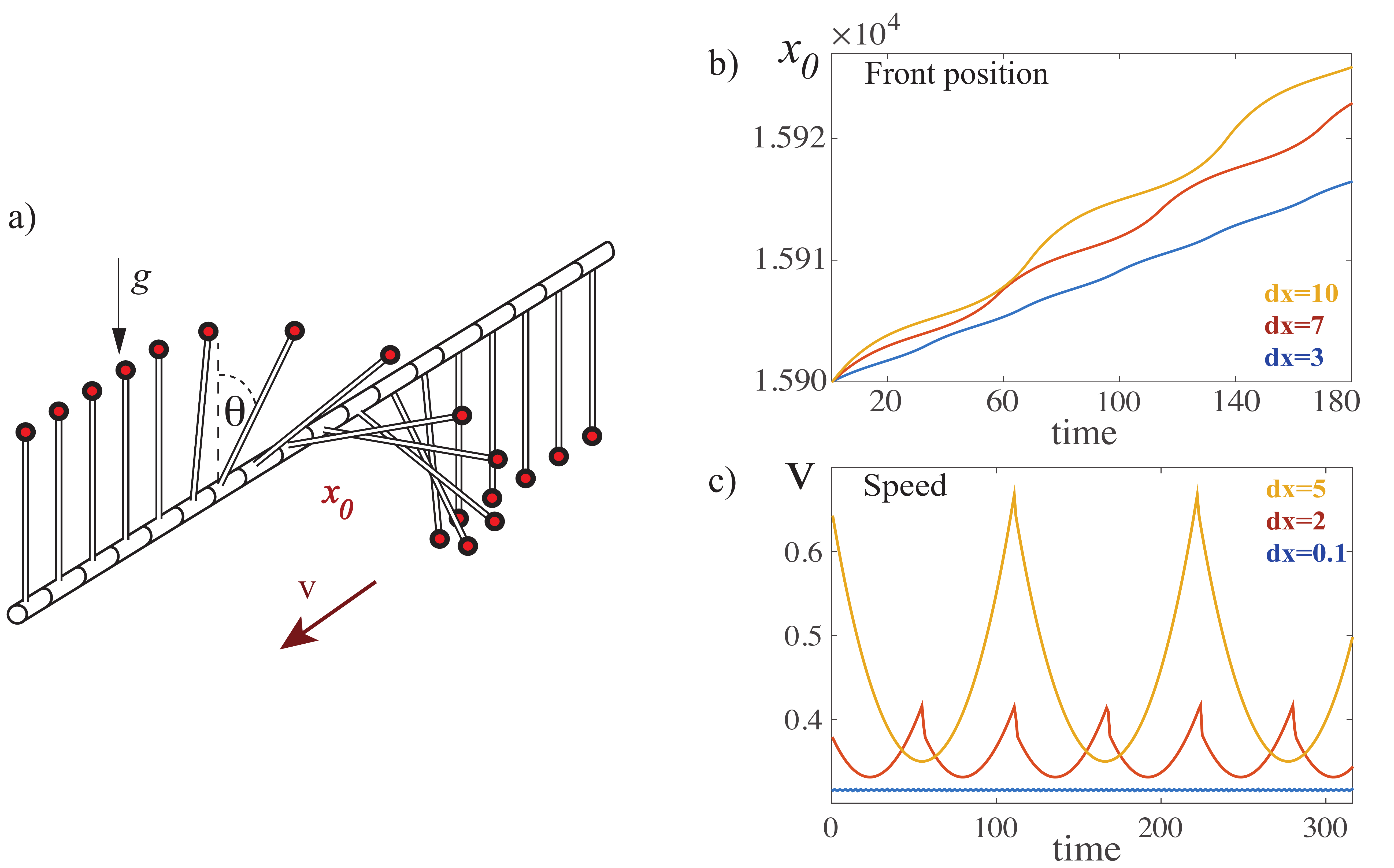} 
\caption{ (color online) Chain of dissipative coupled pendula. 
a) Schematic representation of a chain of dissipative
coupled of pendula. b) Temporal evolution of the front  position $x_0(t)$  with $\omega=1$ 
and $\mu = 20$.
The upper (yellow),  middle  (orange)  and  lower  (blue) lines 
correspond  to $dx = 10$, $dx = 7$, and $dx = 3$, respectively. 
c) Temporal evolution of the front
speed $\dot{x}_0(t)$ with $\omega=1$ and $\mu=6$ . 
The upper  (yellow),  middle  (orange)  and  lower (blue) lines 
correspond  to $dx = 5$, $dx = 2$, and $dx = 0.1$, respectively. }
\label{F-pendula}
\end{figure}

\subsection{Propagation of a $\pi$-kink in a dissipative chain of pendula}

\begin{figure}[h]
\centering
\includegraphics[width=0.95 \columnwidth ]{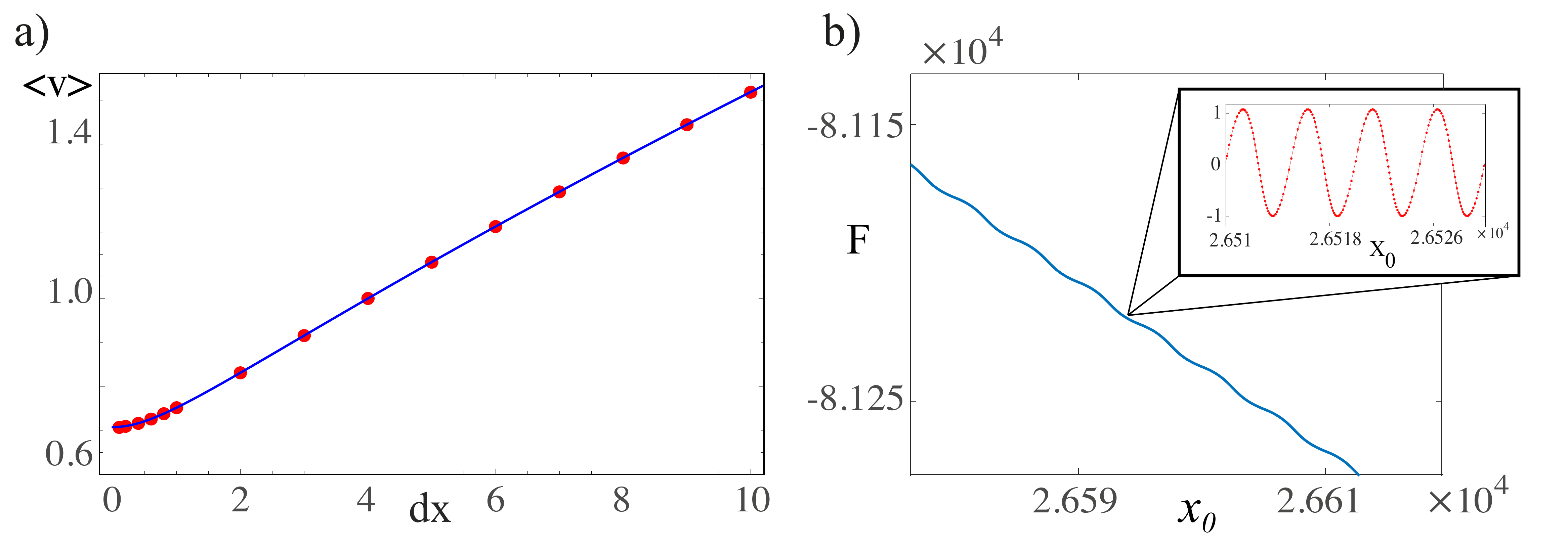}
\caption{Front propagation into an unstable state in discrete chain of dissipative coupled pendula. 
a) Mean front speed as a function of the coupling parameter. Dots (red) are obtained by means of 
numerical simulations of Eq.~(\ref{Eq-ChainPendulum})
with $\omega=1.0$, $dx= 5.0$ and $\mu = 2.0$. The solid line is obtained by using 
the formulas (\ref{Eq-Speedchain}) 
and (\ref{Eq-Beta(dx)pendulums}).
b) Lyapunov functional as function of front position obtained by 
numerical simulations of Eq.~(\ref{Eq-ChainPendulum}) for the same parameters.
Inset: Lyapunov functional computed in the co-mobile system.}
\label{Fig-SpeedLyapunovpendula}
\end{figure}

In the range $\{0, 2 \pi\}$, Eq. (\ref{Eq-ChainPendulum}) has steady 
states $\theta_0=0$ and $\theta_1=\pi $,
which corresponds to the upright  and upside-down position of pendulum, respectively. The
upright (upside-down) position of the pendulum is a  stable (unstable)  equilibrium. Hence, the
chain of dissipative coupled pendula can exhibits domains of upright or upside-down pendula 
as an extended state and a  domain wall or connective front between both domains. 
This solution is usually denominated as  a $\pi$-kink.  
Figure~\ref{F-pendula}a illustrates  a $\pi$-kink solution  of this chain. The position of the domain 
wall, $x_0(t)$, is defined by  a spatial location that interpolates a horizontal pendulum 
(cf. Fig.~\ref{F-pendula}a). 
Due to coupling between pendulums, the  domain wall propagates into the unstable state.
Figures~\ref{F-pendula}b and \ref{F-pendula}c show, respectively, front position and speed 
for different values of discreteness obtained from numerical simulations of equation 
(\ref{Eq-ChainPendulum}). 
Numerical simulations were conducted using finite differences method with Runge-Kutta 
order-4 algorithm
and specular boundary conditions.
Indeed, the speed of propagation of the $\pi$-kink is oscillatory 
with a well defined average speed, $\langle  v \rangle $. Unexpectedly, 
when discreteness $dx$ increases, the mean speed, 
amplitude and frequency of oscillation increases. Note that the oscillations exhibited by the speed
are non-harmonic. Figure~\ref{Fig-SpeedLyapunovpendula}a shows the  mean speed as function of
the discreteness. For large discreteness, the speed increases linearly.

From a numerical solution of a chain of dissipative coupled pendula, we have computed
the Lyapunov functional. Figure~\ref{Fig-SpeedLyapunovpendula}b shows 
evolution of the Lyapunov functional as a function of the front position $x_{0}$. 
As a matter of fact, the Lyapunov functional decreases 
with time in an oscillatory manner.  It is clear from this results that the observed dynamical 
behavior is a consequence 
of the discreteness of the system. 

 To assess the above proposition we shall consider first  a minimal theoretical model---the discrete 
Fisher-Kolmogorov-Petrosvky-Piskunov equation---that contains the main ingredients:  
coexistence between a stable and unstable state in a discrete medium.

\section{The Discrete Fisher-Kolmogorov-Petrosvky-Piskunov model}

Before study the effects of the discreteness in the FKPP-front, we shall establish 
some well-known facts about the front propagation  into unstable states in the continuous case. 

\subsection{ The Fisher-Kolmogorov-Petrosvky-Piskunov model: continuos medium}

The most simple model that present front propagation  into an unstable state is
the Fisher-Kolmogorov-Petrosvky-Piskunov equation,
\begin{equation}
\partial_tu=u (1-u)+\partial_{xx}u,
\label{Eq-FKPPcontinuo}
\end{equation}
where $u(x,t)$ is an order parameter that accounts for an extended transcritical bifurcation.
The above model was used to study the populations dynamics in several contexts \cite{Murray1989},
where the main ingredients are linear growth, nonlinear saturation (logistic nonlinearity),
and Fickian transport process.

In this model, $u=0$ is an unstable
fixed point and $u=1$ is a stable one. If the initial conditions have compact 
support, i.e.,  $u(x,0) = u_{0}(x)$ with
\begin{equation}
u_{0}(x)=\left\{
\begin{array}
[c]{c}
0,\text{ \ \ \ }x<x_{1}\\
f(x),\text{ \ \ \ \ }x_{1}\leq x\leq x_{2}\\
0,\text{ \ \ \ }x<x_{2}%
\end{array}
\right.  \label{Eq-InicialConditions}
\end{equation}
where $f(x)$ is a positive and bounded function, the solution $u(x,t)$ will evolve 
to a two counter propagating  front solutions,
which propagate to speed $v=2$.  In other words, the fronts move with minimum 
speed $v_{min}=2$ \cite{FKP1937,Saarloos2003}. 
This speed was determined by considering traveling solutions 
in the co-mobile dynamical system. After, we perform a linear analysis around 
the unstable state, imposing that the front solutions do not exhibit 
damped spatial oscillations \cite{FKP1937}. Henceforth, we shall denominate this 
method  to determine the minimum speed as {\it FKPP procedure}.   
This result was obtained in the pioneering work of 
Luther \cite{Luther}, in the context of wave propagation in catalytic chemical reaction.
As this minimum speed is determined by means of a linear analysis it is usually 
called {\it linear criterion}. This linear criterion for the determine the minimum speed 
of front propagation into unstable states is valid for  weakly nonlinearity, 
as it is established in the work of Kolmogorov et al. \cite{FKP1937} 
(for more details see Rev. \cite{Saarloos2003} and references therein).
For strong nonlinearity, {\it nonlinear criterion},  the front propagation into unstable 
states also have a minimum speed, however there is no general formulation to determine the value 
of  such speed. 
For gradients equations have been developed a variational method to determine an 
adequate approximation to the minimum 
speed \cite{Benguria}. Fronts whose minimum speed  is determined by the 
linear or nonlinear criteria are usually called {\it pulled} or {\it pushed front}, 
respectively \cite{Saarloos2003}.

As we have mention for different initial conditions, 
the front solution will  strongly depend of the asymptotic behavior of $u(x,0)$ for  $x \to \pm\infty$. 
Considering an  initial conditions of the form $u(x,0) \sim A e^{-kx}$ for $x \to \infty$ 
where $\{k,A\}$ are positive constants. The  front propagates 
as a wave of the form $u(x,t)=u\left(k(x-vt)\right)$. Linearizing  
Eq. (\ref{Eq-FKPPcontinuo}) and considering $u(x,t)\sim A e^{-k(x-vt)}$ for $x \to \infty$, 
after straightforward calculations one can obtain the following relation \cite{Mollison},   
\begin{equation}
v=\frac{1}{k}+k.
\label{Ec: DispersionContinuo}
\end{equation}
Thus, $v$ as function of $k$ is a convex function. 
Minimizing the previous curve with respect to $k$, we obtain the 
critical steepness $k\equiv k_{c}=1$, for which we obtain the minimal 
speed $v(k_c)\equiv v_{min}=2$. 
This method is the {\it asymptotic  process}.
For any other value of the steepness 
$k$ the front propagates with a speed larger than $v_{min}$ ($v(k) \geqslant v_{min}$).
Hence, using the asymptotic shape of the front solution one can determine 
the minimum speed, when the linear criterion is valid.
Note that this procedure can only determine the minimum speed of a pulled front.
It is noteworthy to mention that the asymptotic solution for all co-mobile reference 
systems, i.e., $z=x-vt$ is~\cite{Murray} 
\begin{equation}
\label{Eq:Asymp}
u(z) = \frac{1}{\left(1+ e^{z/v}\right)} + \frac{1}{v^2}\frac{e^{z/v}}{\left(1+e^{z/v}\right)^{2}}
\ln\left[\frac{4e^{z/v}}{\left(1+e^{z/v}\right)^2}\right] + \mathcal{O}\left(\frac{1}{v^{4}}\right)
\end{equation}
where $v\geq  v_{min}=2$. 

\subsection{Front propagation in discrete FKPP model}

Let us consider a simple discrete version of  the Fisher-Kolmogorov-Petrosvky-Piskunov model,  
Eq.(\ref{Eq-FKPPcontinuo}),
\begin{equation}
\dot{u}_i=u_i(1-u_i)+\frac{u_{i+1}-2u_i+u_{i-1}}{dx^2},
\label{Eq-discretFKPP}
\end{equation}
where $u_i(t)$ stands for the population in $i$-th position.
It is assumed that locally the growth is linear,  the saturation is nonlinear, 
and that the population flow is proportional to the population difference of near neighbors.
Note that the dynamics of discrete FKPP model can rewrite in the following form
\begin{equation}
\partial_t u_i=-\frac{\partial {\cal F}}{\partial u_i}
\end{equation}
where the Lyapunov function is defined as
\begin{equation}
{\cal F}=\sum_i\left(-\frac{u_i^2}{2}+\frac{u_i^3}{3}+
\frac{(u_{i+1}-u_i)^2}{2\,dx^2}\right)= \sum_i V_i +\frac{(u_{i+1}-u_i)^2}{2\,dx^2},
\label{Eq-LyapunovFKKP}
\end{equation}
where $V_i$ is the potential. Hence, the dynamics of Eq.~(\ref{Eq-discretFKPP}) 
is characterized by the 
minimization of functional $F$. Indeed, using Eq.~(\ref{Eq-discretFKPP}), one obtains
\begin{equation}
\frac{d{\cal F}}{dt}=\sum_i \frac{\partial {\cal F}}{\partial u_i} \frac{\partial u_i}{\partial t}
=-\sum_i \left(\frac{\partial {\cal F}}{\partial u_i} \right)^2.
\end{equation}

\begin{figure}[t]
\centering
\includegraphics[width=0.9 \columnwidth]{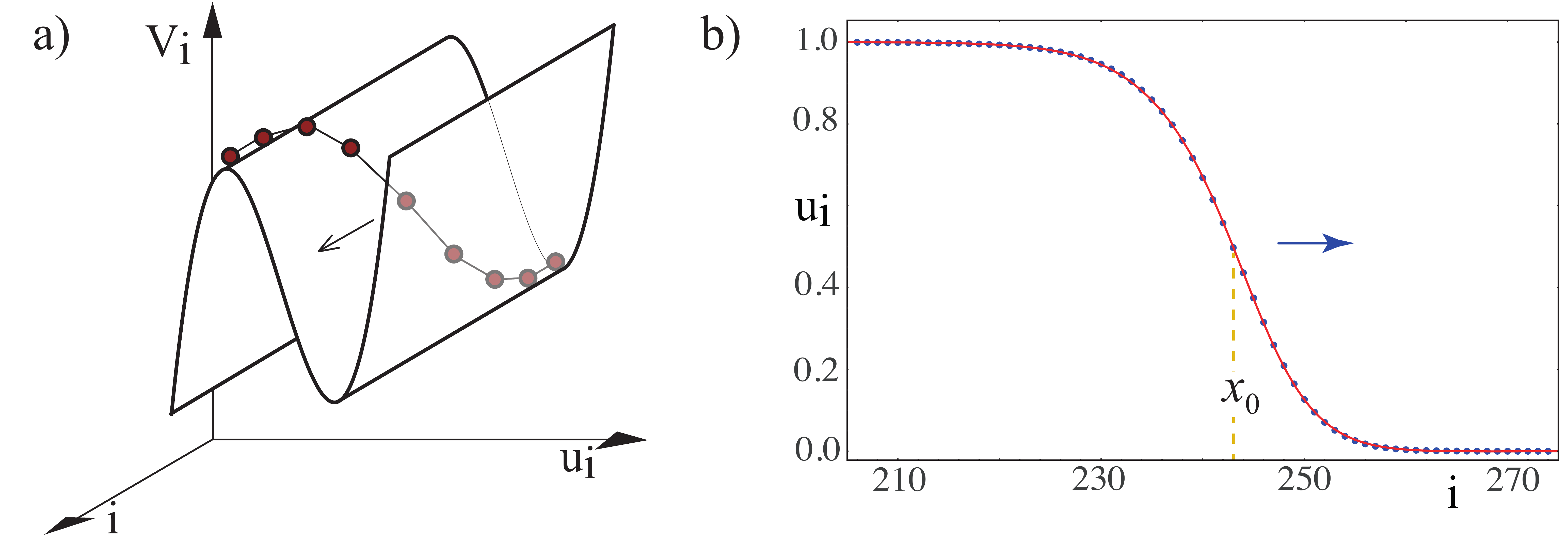} 
\caption{(color online) Front solution of the FKKP model Eq.~(\ref{Eq-discretFKPP}). a)
Schematic representation of potential $V_i$. b) Front solution obtained numerically 
from Eq.~(\ref{Eq-discretFKPP}) (blue dots) and the asymptotic solution 
Eq.(\ref{Eq:Asymp}) (solid line); $x_0$ accounts for the front position.} 
\label{Fig-potencial}
\end{figure}

\begin{figure}[t]
\centering
\includegraphics[width=0.98 \columnwidth]{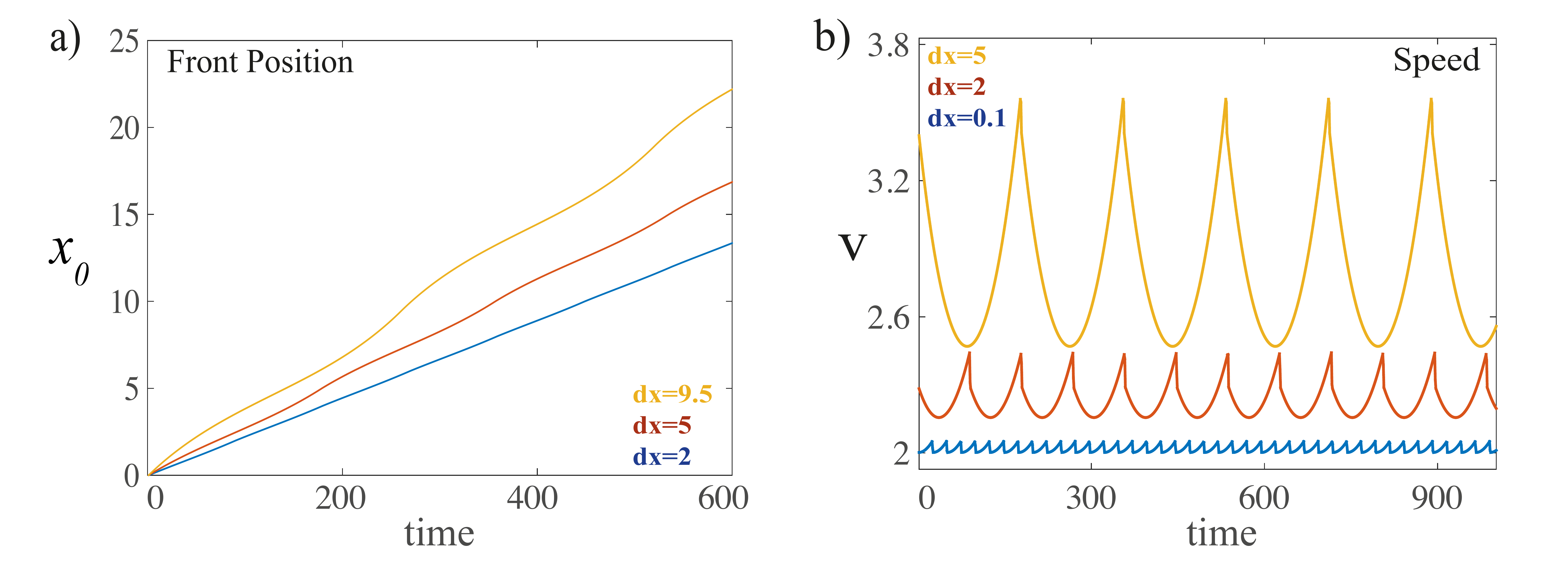}
\caption{(color online) Front propagation into an unstable state in discrete 
FKPP Eq.~(\ref{Eq-discretFKPP}). 
a) Temporal evolution of front  position $x_0(t)$.
The upper (yellow),  middle  (orange)  and  lower  (blue) lines 
correspond  to $dx = 9.5$, $dx = 5$, and $dx = 2$, respectively. 
b) Temporal evolution of  front
speed $\dot{x}_0(t)$. 
The upper  (yellow),  middle  (orange)  and  lower (blue) lines 
correspond  to $dx = 5$, $dx = 2$, and $dx = 0.1$, respectively.}
\label{Fig-FKPPfrente}
\end{figure}
The discrete Fisher-Kolmogorov-Petrosvky-Piskunov Eq.~(\ref{Eq-discretFKPP}) exhibits
front propagation into an unstable state.  In Ref.\cite{Zinner1993}, it
has been established the existence of these solutions, however their oscillatory propagation 
has not been  yet characterized.
Figure~\ref{Fig-potencial} shows a schematic representation of  potential $V_i$ and the front
solution. 
From this figure, one trivially deduces the energy source of the front propagation.

\begin{figure}[ht]
\centering
\includegraphics[width=0.95 \columnwidth]{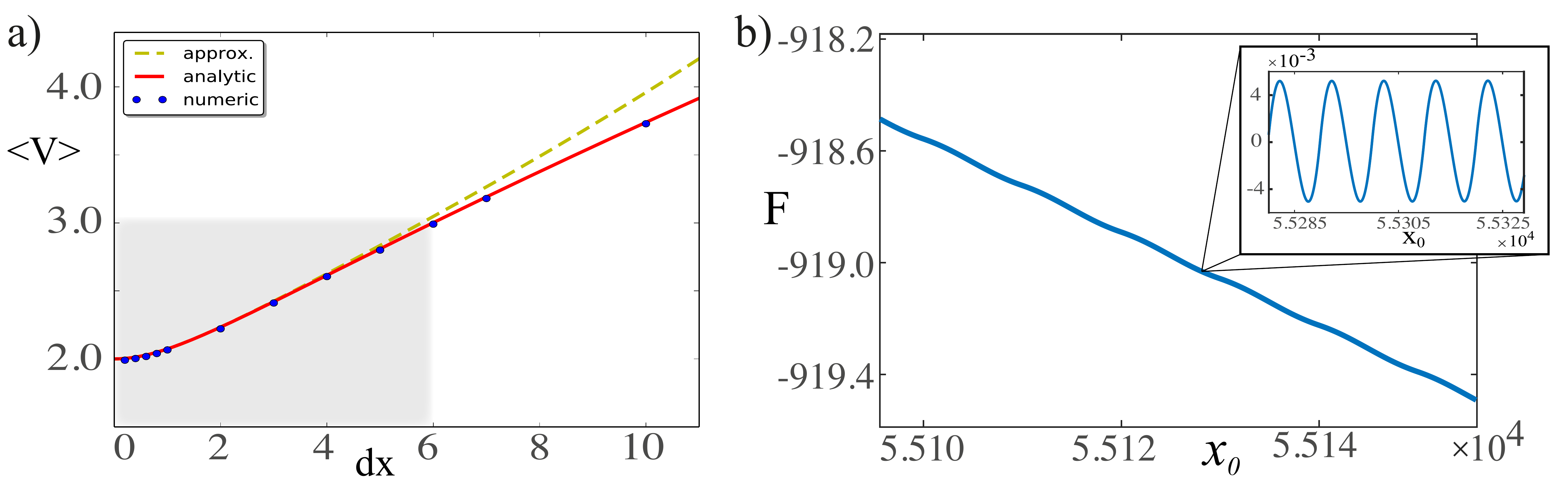}
\caption{(color online) Front propagation into an unstable state in FKPP 
model, equation (\ref{Eq-discretFKPP}).
a) Minimum mean speed as a function of the discreteness $dx$.  Dots (blue) correspond 
to numerical simulations 
of model~(\ref{Eq-discretFKPP}). The solid (red) and dashed (yellow) lines are the 
exact and approximative curve obtained form expressions  
 (\ref{Ec-Speed(dx)implicit}) and (\ref{Eq: MeanSpeed_dx}),
respectively. b) Lyapunov functional as a function of front position
obtained from numerical simulations of Eq.~(\ref{Eq-discretFKPP})
with $dx=10$. Inset: Lyapunov functional computed in the co-mobile system.} 
\label{Fig-Vmean_dx}
\end{figure}

Defining the front position as the spatial position that interpolate the maximum spatial 
gradient, $u (x_0) = 1/2$ (cf. Fig.~\ref{Fig-potencial}b), one can study the front propagation. 
Figure~\ref{Fig-FKPPfrente} shows, respectively, the front position and minimum speed 
for different values of discreteness obtained from numerical simulations of 
model Eq.~(\ref{Eq-discretFKPP}).  We can observe similar dynamical behavior that those 
exhibit by a chain of dissipative coupled 
pendula (cf. Fig.~\ref{F-pendula}), that is,  the front propagates with an oscillatory speed with 
a  given mean speed.  When coupling parameter $dx$ increases, the mean speed, 
amplitude and frequency of oscillations increases. Moreover, the oscillations exhibited by the speed
are non-harmonic type. Figure~\ref{Fig-Vmean_dx}a shows the mean speed as a function of
the coupling parameter. For large $dx$ the speed increases 
linearly.

From a numerical solution of FKPP model Eq.~(\ref{Eq-discretFKPP}), we have computed
the Lyapunov functional (\ref{Eq-LyapunovFKKP}). Figure~\ref{Fig-Vmean_dx}b shows 
evolution of Lyapunov functional as a function of the front position. 
In the inset of Fig.~\ref{Fig-Vmean_dx}b, we show the Lyapunov 
functional in the co-mobile system.
As we can see, Lyapunov functional decreases 
with time {in a} oscillatory manner. 
\begin{figure}[t]
\centering
\includegraphics[width=0.9\columnwidth]{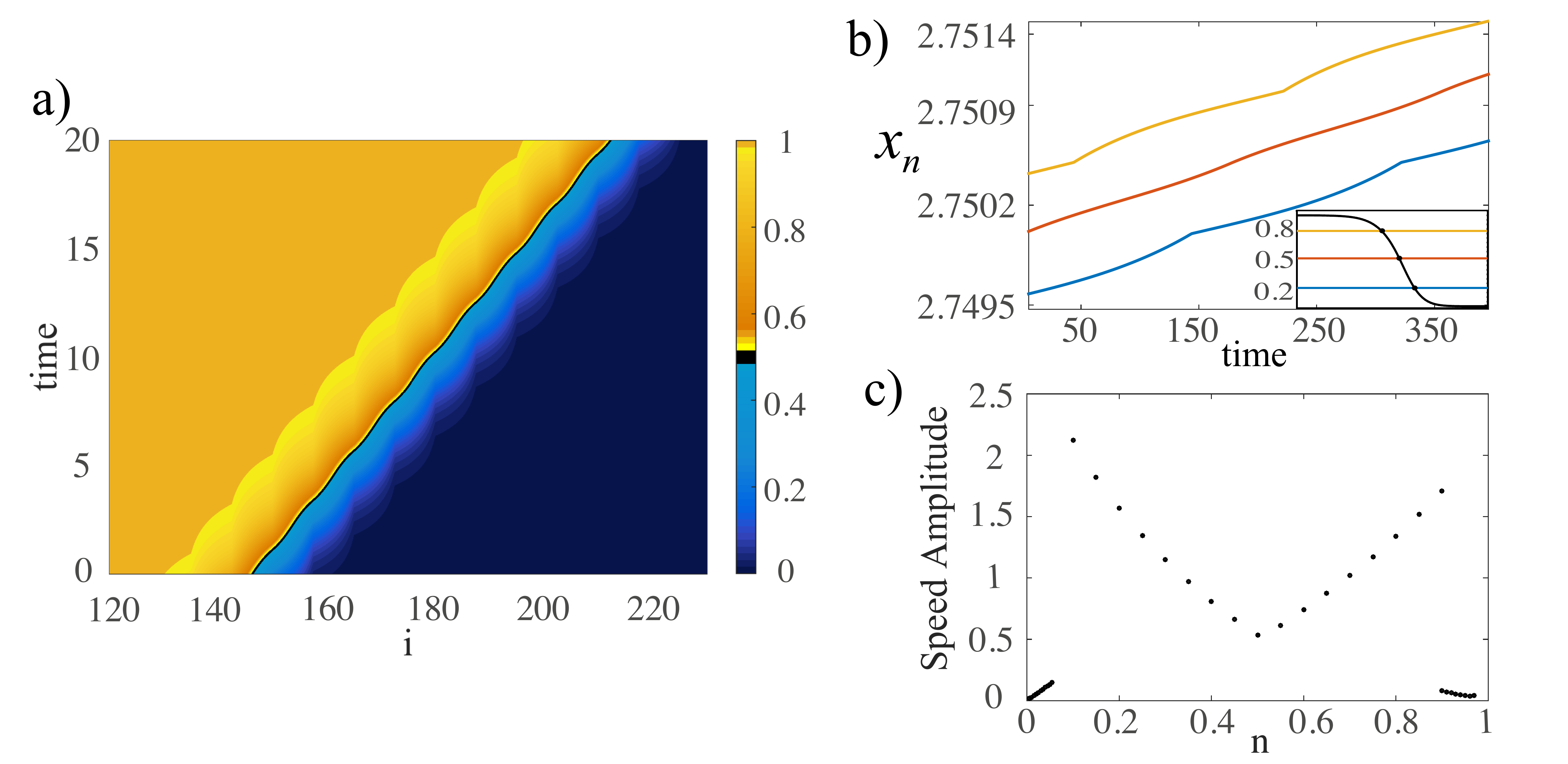}
\caption{Front propagation into an unstable state in discrete FKPP model~(\ref{Eq-discretFKPP}). 
a) Spatiotemporal evolution of the front propagation into an unstable state in
discrete FKPP model~(\ref{Eq-discretFKPP}) with $dx=7.5$.
b) Trajectory of three different  points or cuts: Above  (upper yellow line), 
in (middle red line),  
and below (lower blue line) the front position. {\color{blue}Inset illustrates 
different cuts under consideration.}
c)  Oscillation amplitude of the front speed in different points. } 
\label{Fig-front3D}
\end{figure}

Usually, the study of the front dynamics is reduced to the front position, 
i.e., the dynamical tracking of  point $x_{0}$,  where $u(x_{0}) = 1/2$ 
is the maximum of the spatial gradient. 
This is based on the assumption that the front behaves as a point-like particle. 
Thus,  point $x_{0}$ will gives us enough information about the whole structure dynamics.
Surprisingly, the FKPP front exhibits an extended object behavior: each point 
of the front shows an oscillation dynamics with the same frequency but different amplitude.  
Figure~\ref{Fig-front3D}a shows the  spatiotemporal diagram of the front. 
From this figure,  is easy to infer that the front propagates as an extended object. 
Moreover, the oscillation with respect to  
front position $x_0$ are in anti-phase (see Fig.~\ref{Fig-front3D}b). 
That is, the maximum oscillation of a point to 
the left of the front position coincides with the minimum oscillation of a point to the right. 
To explore the structure of the potential over which the front propagates, 
we have followed different points or "cuts" along the front profile, studying each one separately. 
Figure~\ref{Fig-front3D}c displays the amplitude of oscillation of the front speed  for different cuts. 
From this figure, we conclude that  this amplitude is minimal at the front position, 
increases as one moves away from the front position
and decays to zero  abruptly in the front tails.

\subsection{Theoretical description of the mean speed for the discrete FKPP model}

\begin{figure}[t]
\centering
\includegraphics[width=6.8 cm]{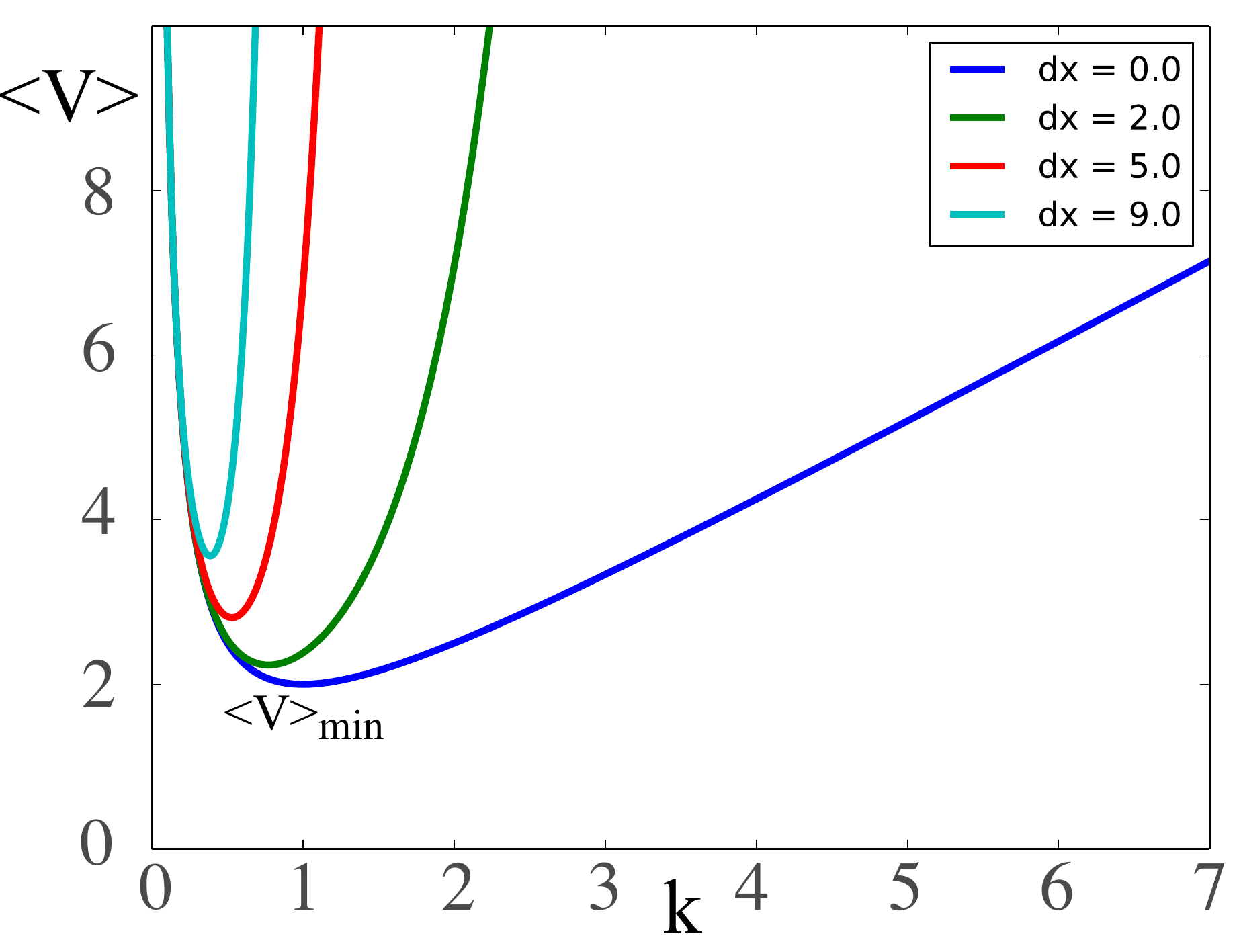}
\caption{Mean speed $\langle v \rangle$ as a function of the steepness parameter $k$ 
for different values of the discretization parameter $dx$, formula~(\ref{Ec: DispersionDiscrete}).
From the lower to upper curve we consider $dx=0, 2,5$ and $9$, respectively. } 
\label{Fig:velocity_k}
\end{figure}

For discrete media, the {\it FKPP procedure} is unsuitable to determine the
minimum speed. Due to there is not a continuos dynamical system 
associated to the co-mobile system inferred for traveling wave solutions.  
To compute the minimal front speed,
we generalize the asymptotic procedure ansatz for the front tail \cite{Mollison}, 
\begin{equation}
\label{Eq-AnsatzFKPP}
u_{i}(t) = e^{\left(\alpha t - 2 i \beta \right)}\left[1+f_{dx; i}^{\omega}(t)\right], \quad \quad \quad i\gg 1,
\end{equation}
with $\alpha \equiv k \langle v\rangle$ and $\beta\equiv  k \,dx/2$ are parameters. 
The index $i\geq 0$ is a positive and large integer number, 
$dx$  is the discretization parameter,  $\langle v \rangle$ is the mean speed of the front, 
and $f_{dx,i}^{\omega}(t)$ is a time periodic function with period $T\equiv 2 \pi/ \omega$, i.e., 
$f_{dx;i}^{\omega}(t) = f_{dx;i}^{\omega}(t +T)$, which accounts for the oscillation of the front
speed at the $i$-th position (cf. Fig.~\ref{Fig-front3D}).  In addition, $f_{dx;i}^{\omega}(t) \to 0$ 
when $ i \to\infty$.  Hence, function  $f_{dx,i}^{\omega}(t)$  takes into account the periodicity 
introduced by the discreteness.  Linearizing  discrete FKKP model~(\ref{Eq-discretFKPP})  
and replacing ansatz~(\ref{Eq-AnsatzFKPP}), 
we get
\begin{equation*}
\begin{split}
\dot{u}_{i} &=\dot{f}_{dx;i}^{\omega} +  \alpha\left[1+f_{dx;i}^{\omega}\right] \\
&= \left[1+f_{dx;i}^{\omega}\right]+ \frac{k^2}{\beta^2}\left[ \left(\sinh^2\left(\beta\right)\right) 
+ \left(\sinh^2\left(\beta\right)\right)f_{dx;i}^{\omega}\right].
\end{split}
\end{equation*}
Integrating this expression in a normalized  period $T$ 
\begin{equation}
\langle\dot{f}_{dx;i}^{\omega} +  \alpha\left[1+f_{dx;i}^{\omega}\right] \rangle
= \langle \left[1+f_{dx;i}^{\omega}\right]+ \frac{k^2}{\beta^2}\left[ \left(\sinh^2\beta\right) 
+ \left(\sinh^2\beta\right)f_{dx;i}^{\omega}\right]\rangle,
\end{equation}
where 
\begin{equation}
\langle g(t) \rangle \equiv \frac{1}{T} \int_0^T g(t) dt,
\end{equation}
we obtain an expression  for the mean speed $\langle v \rangle$
\begin{equation}
\label{Ec: DispersionDiscrete}
\langle v \rangle = \frac{1}{k}+k\left(\frac{\sinh\beta}{\beta}\right)^2.
\end{equation}
with $\langle f_{dx;i}^{\omega}\rangle=
\langle \dot{f}_{dx;i}^{\omega}\rangle =0$, due to $f_{dx;i}^{\omega}(t)$  periodicity. 
This expression accounts for the mean speed as a function of steepness and 
discreteness parameters.
Note that $\langle v \rangle$ tends to expression (\ref{Ec: DispersionContinuo}) 
when $dx\to 0 $ ($\beta\to 0$),
which corresponds to the continuous limit. 
Figure \ref{Fig:velocity_k} shows the mean speed as a function of the parameter $k$ 
for different values of the discretization parameter $dx$. For different values 
of  discretization parameter $dx$, 
$\langle v\rangle$ is a concave function.
We can observe, that the 
minimum speed $\langle v\rangle_{min}$ increases as the discretization 
parameter $dx$ grows. Meanwhile, the critical steepness $k_{c}$ decreases. 
By differentiating the mean speed relation \eqref{Ec: DispersionDiscrete} 
and equating to zero, we obtain an expression for the  discretization parameter, 

\begin{equation}
dx^2=4\sinh\beta_c (2\beta_c \cosh\beta - \sinh\beta_c),
\label{Eq:dx_beta}
\end{equation}
where $\beta_c=k_c dx/2$ and $k_c$ is the critical steepness to obtain the minimum speed. 
Replacing the definition of $\beta_c$ in  above expression, we get
\begin{equation}
dx^2=4\sinh\left(\frac{k_c dx}{2}\right) \left[k_c dx \cosh\left(\frac{k_c dx}{2}\right) 
- \sinh\left(\frac{k_c dx}{2}\right) \right].
\label{Eq-dx_k}
\end{equation}
One can not  explicitly determine the critical steepness 
as fa unction of discretization parameter
$dx$, $k_c(dx)$. Hence, minimum speed 
as a function of $dx$ is a implicit formula 
\begin{equation}
\langle v \rangle_{mim} = \frac{1}{k_c(dx)}+k_c(dx)\left(\frac{\sinh\beta(dx)}{\beta(dx)}\right)^2.
\label{Ec-Speed(dx)implicit}
\end{equation}
The continuos curve in Fig.~\ref{Fig-Vmean_dx}a is the minimal mean speed
as a function of the discretization, expression (\ref{Ec-Speed(dx)implicit}). Numerical 
simulations show quite good agreement with this expression (cf. Fig.~\ref{Fig-Vmean_dx}a).

\begin{figure}[h]
\centering
\includegraphics[width=0.9\columnwidth]{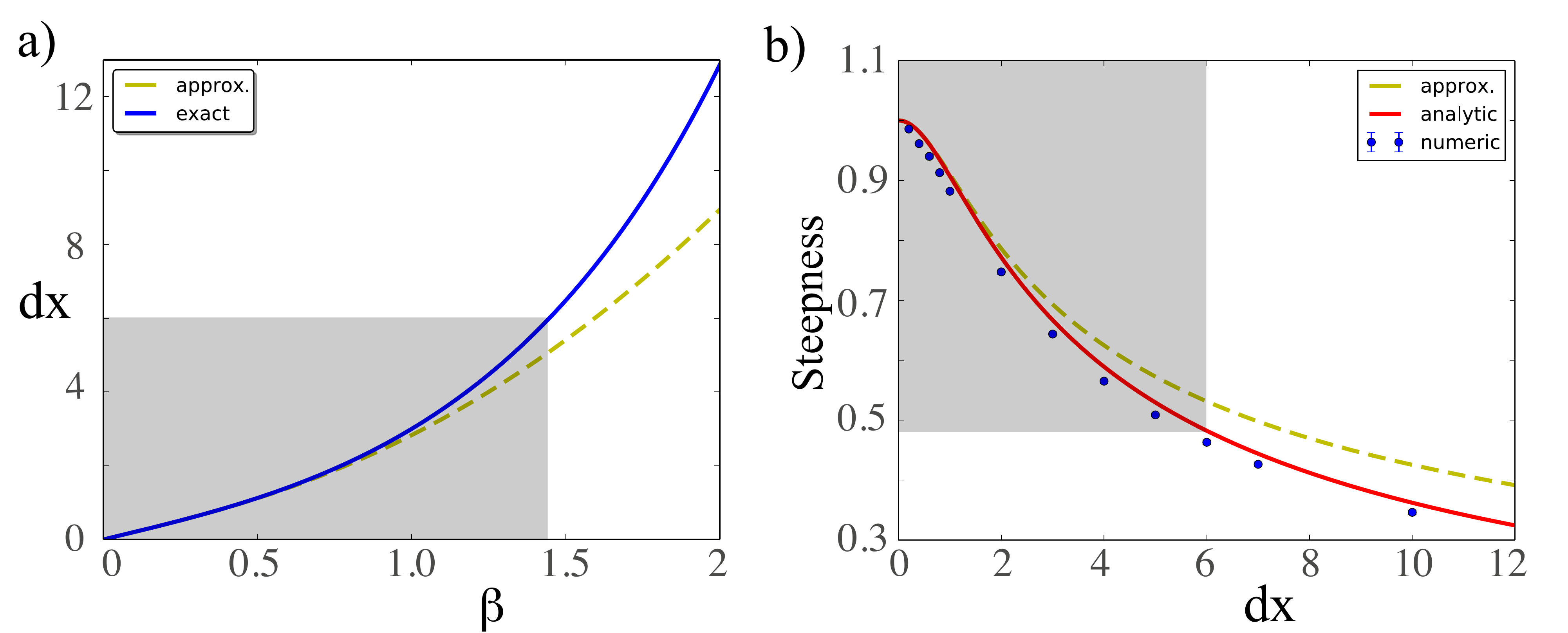}
\caption{(color online) Parameter as function of discretization.
a) Discretization as a function of   $\beta$ parameter. 
Solid (blue) and dashed (yellow) lines are the exact and approximative analytic 
curves  (\ref{Eq-dx_k}) and (\ref{Eq:dx_beta}), respectively.
b) Steepness $k$ as a function of the 
discretization parameter, $dx$. Solid and dashed lines 
are the exact and approximative analytic 
curves (\ref{Eq-dx_k}) and (\ref{Eq: k_dx}), respectively. Dots (blue) are obtained 
by numerical simulations.} 
\label{Fig-dxvsK}
\end{figure}

To have an explicit analytical expression we consider the limit $\beta\to 0$, 
thus  expression (\ref{Eq:dx_beta}) can be simplified to 
\begin{equation}
dx\approx 2\beta_c\sqrt{1+\beta_c^2}.
\label{Eq:dx_beta}
\end{equation} 
From here, we can write the parameter $\beta_c$ 
$$\beta_c\approx \frac{dx}{\sqrt{2\left(1+\sqrt{1+dx^2}\right)}}$$
and the critical steepness  
\begin{equation}
k_c\approx \sqrt{\frac{2}{1+\sqrt{1+dx^2}}}
\label{Eq: k_dx}
\end{equation}
as a function of the discretization parameter $dx$. Figure \ref{Fig-dxvsK} shows  
the discretization parameter $dx$ as a function of $\beta$ (Eq. \eqref{Eq:dx_beta}). 
The shadow area illustrates the limit where the approximation $\beta\to 0$ is valid. 
Likewise, figure~(\ref{Fig-dxvsK}) shows the steepness $k$ as a function of the 
parameter $dx$. From both, we can infer that expressions \eqref{Eq:dx_beta} 
and \eqref{Eq: k_dx} are valid in a wide range of the parameter $dx$. Therefore, 
in a good approximation the mean speed $\langle v \rangle$ can take the form,
\begin{equation}
v\approx \sqrt{\frac{1+\sqrt{1+dx^2}}{2}}\left[1+\frac{4}{dx^2}
\sinh^2\left(\frac{dx}{\sqrt{2\left(1+\sqrt{1+dx^2}\right)}}\right)\right].
\label{Eq: MeanSpeed_dx}
\end{equation}

Figure \ref{Fig-Vmean_dx} shows the mean speed as a function of the 
discretization parameter $dx$. Up to a value of $dx=6.0$,  expression  \eqref{Eq: MeanSpeed_dx} 
is an adequate approximation. We observe a good accordance between the analytic expression  and the  
mean speed obtained by numerical simulations. 

In brief, {\it the asymptotic procedure} allows an adequate characterization of the  average features
of  front  propagation into unstable states in discrete media.
In the next section, we shall apply this procedure to characterize the mean
properties of front propagation into unstable state in a 
chain of dissipative coupled pendula.

\subsection{Theoretical description of the mean speed for the Chain of dissipative coupled pendula}

For the chain of dissipative coupled pendula, Eq.~(\ref{Eq-ChainPendulum}), 
the unstable state correspond to
$\theta_i=\pi/2$. Considering  the asymptotic ansatz for the  front tail 
around this state, we get,
\begin{equation}
\label{Eq:AnsatzFKPP}
\theta_{i}(t) = \frac{\pi}{2}+A_0 e^{\left(\alpha t - 2 i \beta \right)}\left[1+f_{dx;i}^{\omega}(t)\right],
\end{equation}
where $A_0$ is a constant that characterizes the shape of the front tail, 
$\alpha \equiv k \langle v\rangle$ and $\beta\equiv  k \,dx/2$ are parameters. 
$f_{dx;i}^{\omega}(t)$ is  a periodic function of frequency $\omega$ in  $i$-th position
of the chain that describes the
oscillatory behavior of the speed. Introducing the above ansatz in Eq.~(\ref{Eq-ChainPendulum}) and 
taking into account only the linear leading terms, we obtain, 
\begin{equation*}
\begin{split}
\ddot{\theta}_{i} &=\alpha^2\left[1+f_{dx;i}^{\omega}(t)\right] + 2\alpha \dot{f}_{dx;i}^{\omega}(t) 
+  \ddot{f}_{dx;i}^{\omega}(t) \\
&= \omega^2\left[1+f_{dx;i}^{\omega}(t)\right] -\mu\left[ \alpha \left[1+f_{dx;i}^{\omega}(t)\right]  
+ \dot{f}_{dx;i}^{\omega}(t)\right] + \frac{4}{dx^2}\sinh^2(\beta) \left[1+f_{dx;i}^{\omega}(t)\right]
\end{split}
\end{equation*}
Integrating this expression in a normalized  period $T=2\pi/\omega$, and considering
$\langle f_{dx;i}^{\omega}(t) \rangle=\langle \dot f_{dx;i}^{\omega}(t) \rangle
=\langle \ddot f_{dx;i}^{\omega}(t) \rangle=0$, after straightforward calculations, we obtain
\begin{equation}
\begin{split}
\alpha^2 = \omega^2 -\mu\alpha  + \frac{4}{dx^2}\sinh^2(\beta).
\end{split}
\end{equation}
Substituting the definition of $\alpha$, the mean speed reads
\begin{equation}
\label{Ec: DispersionDiscretePendula}
\langle v \rangle = -\frac{\mu}{k} + \frac{1}{k}\sqrt{\mu^2 + \omega^{2} 
+ \frac{k^2}{\beta^2} \sinh^2(\beta)},
\end{equation}
and replacing $k= 2\beta/dx$, 
\begin{equation}
\langle v \rangle = -\frac{\mu}{2\beta }dx + \frac{1}{2\beta}\sqrt{dx^2\left(\mu^2 
+ \omega^{2}\right) + 4 \sinh^2(\beta)}
\label{Eq-Speedchain}
\end{equation}
The above expression accounts for front speed as a function of the steepness.
In order to deduce the minimal front speed, we differentiate  the above speed with 
respect to $\beta$
\begin{equation}
\begin{split}
&\omega^2\left(\mu^2 + \omega^2\right)dx^4 + 4\left[ \left(\mu^2 
+ 2\omega^2\right)\sinh^2(\beta) - 2\left(\mu^2 + 
\omega^2\right)\beta\sinh(\beta)\cosh(\beta)\right]dx^2\\ 
&+ 16\sinh^{2}(\beta)\left[\sinh(\beta) - \beta\cosh(\beta)\right]^2=0.
\end{split}
\label{Eq-Beta(dx)pendulums}
\end{equation}
This expression gives us a relation between the critical  steepness $k_c$ 
and the coupling parameter $dx$.
An explicit expression $k_c(dx)$ cannot be derived.
Using expression (\ref{Eq-Beta(dx)pendulums}) in  formula (\ref{Eq-Speedchain}), 
we obtain the minimal 
front speed for the chain of dissipative coupled pendula, Eq~(\ref{Eq-ChainPendulum}).
Note that this analytical results has quite fair agreement with the numerical simulations
as it is shown in Fig.~\ref{Fig-SpeedLyapunovpendula}a. Therefore, the asymptotic procedure 
is a suitable method to characterize the mean properties of  front propagation.

\section{Effective continuous model: oscillatory properties of front propagation}

Due to the complexity of discrete dissipative systems, to obtain 
analytical results is a daunting task.
In order to figure out the oscillatory behavior of the front, we shall consider a similar strategy 
to that used in Ref. \cite{ClercEliasRojas}, which is based on considering an effective 
continuous equation that accounts for the dynamics of the discrete system.
The benefit of this approach  is that analytical calculations are accessible.

\begin{figure}[h]
\centering
\includegraphics[width=0.95\columnwidth]{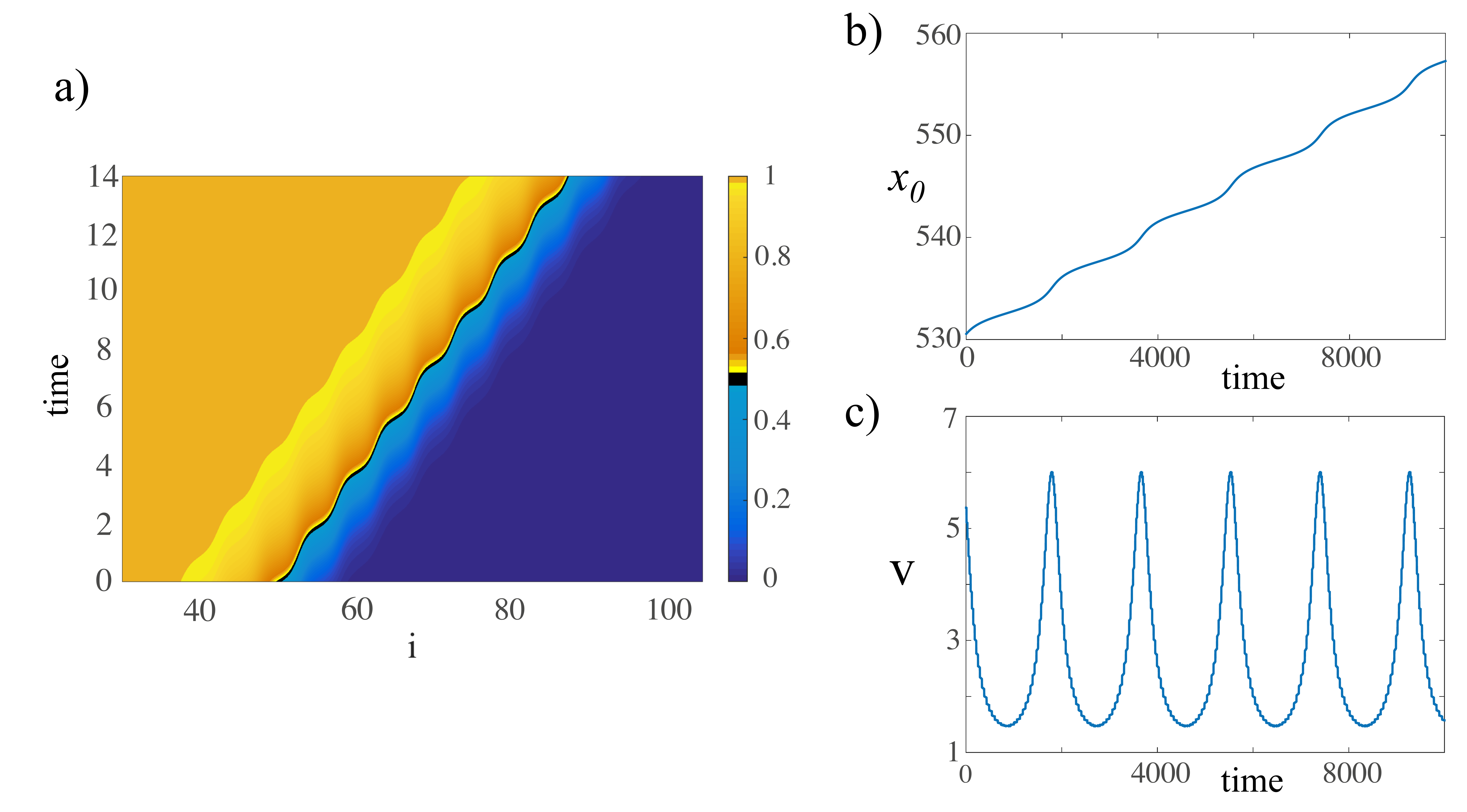}
\caption{(color online) Front propagation into an unstable state in  FKPP Eq.~(\ref{Eq-EffectiveFKPP})
with a harmonic generalized  Peierls-Nabarro potential, $\Gamma_{dx}(x)=A \cos (2\pi x/ dx)$
with $A=0.06$, and $dx=5.0$. The numerical discretization parameter of 
{\color{blue} the finite differences method} is $0.1$. 
a) Spatiotemporal evolution of the front propagation into unstable state. 
b) Temporal evolution of the front position  and  c) minimum speed.} 
\label{Fig-SpeedEffectiveFKPP}
\end{figure}

\subsection{Generalized  Peierls-Nabarro potential}

Let us consider the continuos order parameter $u(x,t)$, which satisfies 
\begin{equation}
\partial_tu=-\frac{\delta {\cal F}}{\delta u},
\end{equation}
where the Lyapunov functional has the form
\begin{equation}
{\cal F}=\int \left(-\frac{u^2}{2}+\frac{u^3}{3}+\frac{(\partial_xu)^2}{2}
+(\partial_xu)^2\Gamma_{dx}(x)\right)dx,
\end{equation}
$\Gamma_{dx}(x)$ is a spatial periodic function with $dx$ period, 
$\Gamma_{dx}(x+dx)=\Gamma_{dx}(x)$.
This function accounts for the discreteness of the system.
The last term of the free energy is a generalization of the Peierls-Nabarro potential.
An effective potential has been used to explain the dynamics 
of defects position such as dislocations in condensed matter physics 
or dynamics of the position of kink or fronts 
(see textbook \cite{FrenkelKontorova} and reference therein). Here, 
we consider an effective equation for the entire field $u(x,t)$, which reads
\begin{equation}
\partial_tu= u (1-u) + D\partial_{xx}u +
2\Gamma_{dx}(x) \partial_{xx}u +
2\Gamma'_{dx}(x) \partial_xu.
\label{Eq-EffectiveFKPP}	
\end{equation}
This equation is a populations dynamical model  
with linear growth, nonlinear saturation,  inhomogeneous diffusion  and drift force. 
Numerical simulations with a harmonic potential  $\Gamma_{dx}$ exhibit front solutions.
It is important to note that for these numerical simulations, 
we have discretized the Laplacian and gradient of $u$ to first neighbors considering a small $dx$.
For which the discreteness  effects are negligible.
Figura~\ref{Fig-SpeedEffectiveFKPP} shows the spatiotemporal diagram
of the front into an unstable state of the effective FKPP, Eq.~(\ref{Eq-EffectiveFKPP}),
with a harmonic generalized  Peierls-Nabarro potential. Its trajectory and speed 
are also illustrated. We can observe that  the numerical simulations
of the effective FKPP, Eq.~(\ref{Eq-EffectiveFKPP}), and discrete  FKPP, Eq.~(\ref{Eq-discretFKPP}),
have similar qualitative dynamical behaviors.

To understand better the generalized  Peierls-Nabarro potential, Figure~\ref{Fig-PNEffectiveFKPP}a 
shows the effective force for the harmonic case and the amplitude speed for the 
effective FKPP Eq.~(\ref{Eq-EffectiveFKPP}). From this figure, we infer that the effective 
force, $f\equiv2\Gamma_{dx}(x) \partial_{xx}u + 2\Gamma'_{dx}(x) \partial_xu$, 
has an oscillatory structure concentrated in the region where the front displays larger spatial variations.
Moreover, we observe that the structure of the amplitude  of the speed is similar to 
that observed in the discrete case (cf. Figs.~\ref{Fig-PNEffectiveFKPP}b and  \ref{Fig-front3D}c).

\begin{figure}[h]
\centering
\includegraphics[width=0.95\columnwidth]{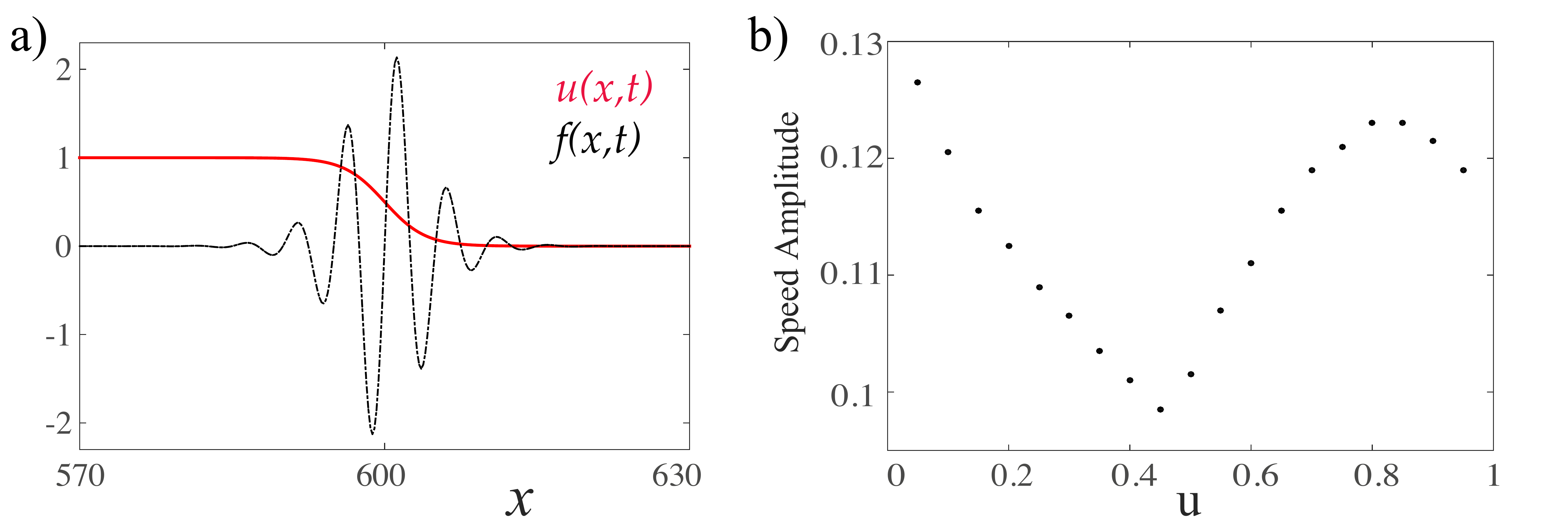}
\caption{The generalized  Peierls-Nabarro force for a harmonic case, 
$\Gamma_{dx}(x)=A \cos (2\pi x/ dx)$. 
a) Front solution and effective force $f\equiv2\Gamma_{dx}(x) \partial_{xx}u 
+ 2\Gamma'_{dx}(x) \partial_xu$. 
b) Amplitude of the speed for  the effective FKPP Eq.~(\ref{Eq-EffectiveFKPP}) 
with $D=1.97$, $A=0.03$, and $dx=5.0$. The numerical {\color{blue}discretization parameter 
of the finite differences method is $0.1$.}} 
\label{Fig-PNEffectiveFKPP}
\end{figure}

\subsection{Dynamics of front position}

The equilibria are not affected by the presence of the periodical extra terms, effective force.
In the continuous limit,  $dx\to 0 $ and $\Gamma_{dx}(x) \to 0$ one recovers the 
Fisher-Kolmogorov-Petrosvky-Piskunov model Eq.~(\ref{Eq-FKPPcontinuo}).
Then for $dx \ll 1$, the last two terms of Eq.~(\ref{Eq-EffectiveFKPP}) are perturbative.
We shall analyze this region of parameters, where we can obtain analytical results.

The Fisher-Kolmogorov-Petrosvky-Piskunov model Eq.~(\ref{Eq-FKPPcontinuo})
has front solutions of the form $u_{FKPP}(x-vt-p)$, where $p$ is a constant that accounts 
for the front position and $v$ the front speed. Analytical expressions of this solution 
are unknown, however solutions in the form of perturbative series are available \cite{Murray1989}.
Considering the following ansatz for small discreteness ($dx \ll 1$)
 \begin{equation}
 u(x,t)=u_{FKPP}\left(x-vt-p(t)\right)+w(x-vt-p(t),p(t)),
\end{equation} 
where  front position is promoted to a temporal function, $p(t)$ and 
$w$ is a corrective function on the order of the perturbative force. 
Introducing the above ansatz in Eq.~(\ref{Eq-EffectiveFKPP}) and linearizing  in   $w$,
after straightforward calculations, we obtain
\begin{equation}
{\cal L}w=-\dot{p}(t)\partial_\xi u_{FKPP}-2 \Gamma_{dx}(x) \partial_{\xi\xi}u_{FKPP}
-2 \Gamma'_{dx}(x) \partial_{\xi}u_{FKPP},
\label{Eq-LinearW}
\end{equation}
where ${\cal L}\equiv \partial_{\xi\xi}+v\partial_\xi+1-2 u_{FKPP}(\xi)$ 
is a linear operator and $\xi =x-vt-p$ 
is the coordinate in the co-mobile system. Considering the inner product 
\begin{equation}
\left<f |g\right> = \int_{-L}^Lf(\xi)g(\xi)d\xi,
\end{equation}
where $2L$ is the  system size. 
In order to solve the linear Eq.~(\ref{Eq-LinearW}), we apply
the Fredholm alternative or solvability condition \cite{Pismen}, and obtain
\begin{equation}
\dot{p}(t) = -2\frac{\left< \Gamma_{dx}(\xi+vt+p) \partial_{\xi\xi}u_{FKPP}| \psi\right>}
{\left<\partial_\xi u_{FKPP} | \psi\right>}-2\frac{\left< \Gamma'_{dx}(\xi+vt+p) 
\partial_{\xi}u_{FKPP}| \psi\right>}
{\left<\partial_\xi u_{FKPP} | \psi\right>},
\label{Eq-FrontPosition}
\end{equation}
where $\psi(\xi)$ is an element of kernel of adjoint of $\cal L$, 
 ${\cal L^\dag}\equiv \partial_{\xi\xi}-v\partial_\xi+1-2 u_{FKPP}(\xi)$,
that is ${\cal L}^\dag\psi=0$. 
The $\psi$ function is unknown analytically, 
however the asymptotic  behavior of this function are characterized to diverges exponentially with the 
the same exponent  that $u_{FKPP}$ converges to their equilibria.
Therefore the above integrals diverge proportional to $L$, however the ratio is well defined.

To understand the dynamics described by the above equation for simplicity we shall consider the 
generalized Peierls-Nabarro potential for a harmonic case, that is,
\begin{equation}
\label{Eq:Gamma_x}
\Gamma_{dx}(x)=\gamma(x)\equiv A \cos	\left(\frac{2 \pi x}{dx}\right),
\end{equation}
Replacing this expression in Eq.~(\ref{Eq-FrontPosition}), after straightforward calculations, we obtain
\begin{equation}
\dot{p}(t) = \sqrt{K_1^2+K_2^2} \cos\left(\frac{2 \pi }{dx} (p-vt)+\phi_0\right),
\label{Eq-FrontPositionArmonic}
\end{equation} 
with
\begin{eqnarray}
K_1 & = & A\frac{\left< \cos\left(\frac{2 \pi\xi}{dx} \right)\partial_{\xi\xi} u_{FKPP}(\xi)
-\frac{2 \pi\xi}{dx} \sin\left(\frac{2 \pi\xi}{dx} \right)\partial_{\xi} u_{FKPP}| \psi(\xi) \right>}
{\left<\partial_\xi u_{FKPP}| \psi(\xi) \right>}, 
\nonumber \\
K_2 & = &- A\frac{\left< \sin\left(\frac{2 \pi\xi}{dx} \right)\partial_{\xi\xi} u_{FKPP}(\xi)
+\frac{2 \pi\xi}{dx} \cos\left(\frac{2 \pi\xi}{dx} \right)\partial_{\xi} u_{FKPP}| \psi(\xi) \right>}
{\left<\partial_\xi u_{FKPP}| \psi(\xi) \right>},
\nonumber \\
\tan(\phi_0) &=& \frac{K_1}{K_2}.
\end{eqnarray}
Therefore, the front position propagates in a oscillatory manner. Notice that the Peierls-Nabarro potential
 propagates together with the front. Figure \ref{Fig-PNEffectiveFKPP} shows fitting curve for $\dot{p}$ using solution of  \eqref{Eq-FrontPositionArmonic} if $\Gamma_{dx}(x)$ is given by  \eqref{Eq:Gamma_x}.
 We can see  that the analytical result is in good agreement with the observed dynamics.

 \begin{figure}[h]
\centering
\includegraphics[width=0.55\columnwidth]{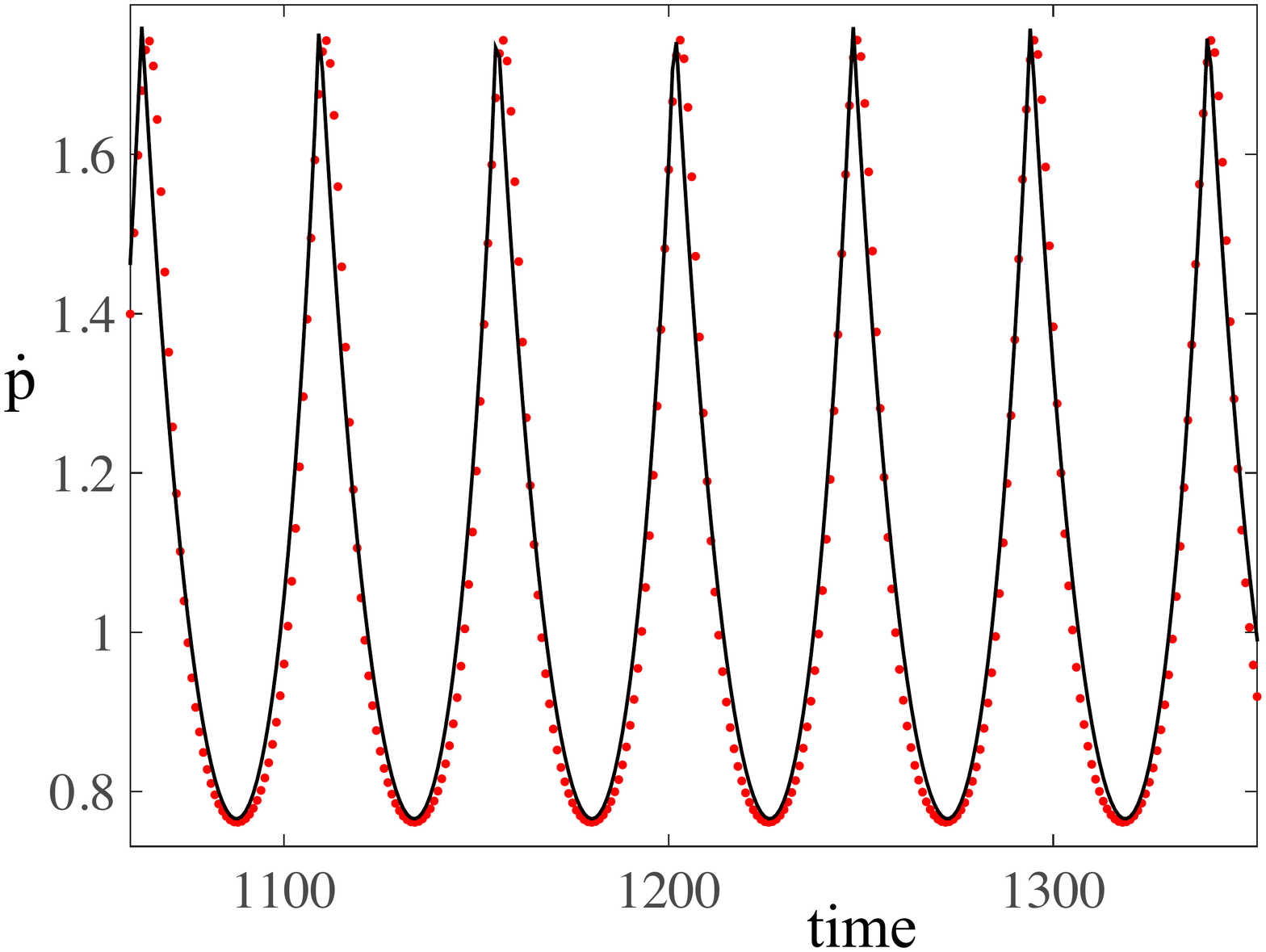}
\caption{Fitting curve for $\dot{p}$ given by the expression
$\ \dot{p} = \frac{a \sec^2(bx+c)}{1+\left(d \tan(bx+c) - e \right)^2} + f$ with 
$a = 0.57$, $b = 0.0682$, $c = 0.6157$, $d = 1$, $e = 0.8$, and  $f = 0.5$}
\label{Fig-PNEffectiveFKPP}
\end{figure}

\section{Conclusions and remarks}

We have studied, theoretically and numerically, the front propagation
into unstable states in discrete dissipative systems. Based on a paradigmatic model of 
coupled chain of oscillators (the dissipative Frenkel-Kontorova model) and population dynamic model
(the discrete and effective Fisher-Kolmogorov-Petrosvky-Piskunov model), we have determined 
analytically the mean speed of  FKPP fronts when nonlinearities are weak. 
Numerically we have characterized the oscillatory front propagation. 
Likewise, we have revealed that different parts of the front oscillate with the same frequency 
but with different amplitude. To describe 
this phenomenon, we have generalized the notion of the  Peierls-Nabarro potential,
which  allows us to have an effective continuous description of discreteness effect.

The analysis presented only is valid for weak nonlinearity, where linear criterium is valid.
The characterization of  pushed front  in local coupling dissipative systems is an open 
and relevant question.
 Propagation fronts in two dimensions is affected by the curvature of the interface, which can 
 increase o decrease the speed of the propagating interface.
Study of front propagation into unstable states in these contexts are in progress.

\section*{Acknowledgments}

M.G.C., M.A.G-N., and R.G.R thank for the financial support of FONDECYT projects 1150507, 11130450, 
and 1130622, respectively. K.A-B. was supported by CONICYT, scholarship Beca de
Doctorado Nacional No.21140668.


\begin{thebibliography}{00}
\bibitem{FKP1937} Kolmogoroff A, Petrovsky I, Piscounoff N. Study of 
the diffusion equation with growth of the quantity of matter and its application 
to a biology problem. Bulletin de l'universit\'e d'\'etat \`a Moscou 
Ser int Section A;1:1.
\end{thebibliography}
\end{document}